\documentclass[11pt]{article}

\usepackage[preprint]{acl}

\usepackage{times}
\usepackage{latexsym}
\usepackage{graphicx}
\usepackage{booktabs}
\usepackage{amsmath}
\usepackage{amssymb}
\usepackage[T1]{fontenc}

\usepackage[utf8]{inputenc}

\usepackage{microtype}

\usepackage{inconsolata}

\usepackage{graphicx}

\title{Mixture-of-Experts Language Models Can Be Strong and Efficient Retrievers}

\author{
  \textbf{Anubhav Shrestha\textsuperscript{1}}\thanks{Corresponding author: \href{mailto:anubhav.shrestha@nyu.edu}{anubhav.shrestha@nyu.edu}},
  \textbf{Safal Shrestha\textsuperscript{1}},
  \textbf{Minwu Kim\textsuperscript{1}},
  \textbf{Torsten Suel\textsuperscript{2}},
  \textbf{Keith Ross\textsuperscript{1}}
\\
\\
  \textsuperscript{1}New York University Abu Dhabi \quad
  \textsuperscript{2}New York University
}

\begin{document}
\setcounter{footnote}{1}
  \maketitle 
\begin{abstract}
Recent work has shown that fine-tuning decoder-only large language models
(LLMs) for retrieval yields strong first-stage retrievers, with effectiveness
improving as backbones grow in size. However, every query and document must
pass through the full model, so encoding cost increases with model size.
Mixture-of-Experts (MoE) LLMs activate only a subset of parameters per token
and are widely used to scale generative models, yet remain underexplored as
retrievers. We systematically study MoE backbones for retrieval by training
MoE and dense LLMs from several families using the same procedure, evaluating
them across diverse datasets, and measuring query encoding time under the same
serving configuration. We show that MoE retrievers outperform dense retrievers
with comparable active parameter counts by up to 3.0 nDCG@10 points on BEIR.
One of our strongest MoE retrievers matches an 8B dense retriever with 59\% fewer active parameters and 18\% lower query encoding time. We further show that the
number of experts used for query encoding can be reduced without retraining or
re-indexing, retaining more than 99\% of retrieval effectiveness while
reducing query encoding time by up to 26\%. Recent rerankers provide only
modest additional gains over strong MoE first stages, which often match or
exceed the reranked configurations we evaluate. Together, these results show
that MoE LLMs can be strong and efficient first-stage retrievers.

\end{abstract}

\section{Introduction}

\begin{figure}[t]
  \centering
  \includegraphics[width=\columnwidth]{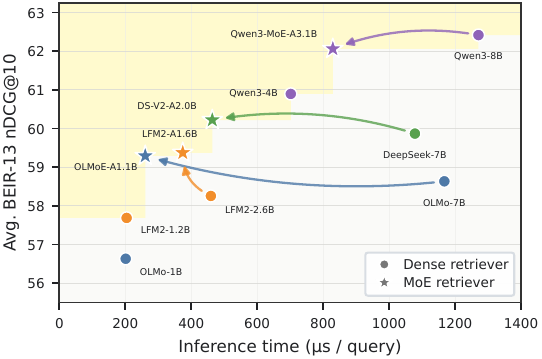}
  \caption{Effectiveness--efficiency trade-off for MoE ($\bigstar$) and dense ($\bullet$) LLM retrievers. Each point shows a model's average BEIR-13 nDCG@10 against its query inference time. Arrows
  connect each MoE retriever to the larger dense model of its family: MoE retrievers reach comparable or higher effectiveness than dense models that have 1.6 to 6.4 times more active parameters while
  reducing query inference time by 19\% to 78\%. MoE retrievers consistently lie near the Pareto frontier (shaded).}
  \label{fig:frontier}
\end{figure}

Information retrieval (IR) identifies documents relevant to a query and is a core component of web search, open-domain question answering, fact verification, and retrieval-augmented generation \cite{bajaj2016msmarco,chen2017drqa,thorne2018fever,lewis2020rag}. Sparse retrievers such as BM25 rank documents by term overlap, while learned sparse methods such as SPLADE expand and reweight terms to reduce vocabulary mismatch \cite{robertson2009bm25,formal2021splade}. Dense retrievers instead encode queries and documents into vector representations and rank them by similarity \cite{karpukhin2020dpr}, and have become a strong approach to first-stage retrieval \cite{izacard2022contriever,wang2022e5,xiao2024cpack}.

Decoder-only LLMs have recently become strong dense-retrieval backbones. RepLLaMA showed that these models can be fine-tuned as bi-encoders \cite{ma2024repllama}, and subsequent work has improved the training data, objectives, and architectures used to adapt LLMs for retrieval \cite{wang2024e5mistral,behnamghader2024llm2vec,lee2025nvembed,zhang2025qwen3embedding}. Across several model families, larger backbones generally yield stronger retrievers \cite{ni2022gtr,zhang2025qwen3embedding,xu2026laconic}. This scaling, however, increases encoding cost. Documents can be encoded offline, but every query must still pass through the full backbone \cite{ma2025lightretriever}, making query encoding cost an increasingly important concern for LLM-based retrieval \cite{abdallah2026worth,lei2025effir}.


MoE architectures offer a natural alternative to dense scaling. By routing each token to a small subset of experts, they allow total model capacity to grow without activating all parameters for every token \cite{lepikhin2021gshard,fedus2022switch}. MoE backbones now appear in several prominent open-weight LLM families, including Mixtral, DeepSeek, OLMoE, and Qwen3 \cite{jiang2024mixtral,deepseek2024v2,muennighoff2025olmoe,yang2025qwen3}. However, public evidence for using recent pretrained MoE LLM backbones as first-stage retrievers remains limited. Nomic Embed v2 trains a purpose-built, BERT-scale MoE encoder specifically for text embedding \cite{nussbaum2025nomic}, while GritLM evaluates a single Mixtral-based model under a joint generative and embedding objective and reports lower retrieval effectiveness than its dense counterpart \cite{muennighoff2025gritlm}. Proprietary MoE embedding models have also been released, but provide few public details about their architectures or measured efficiency \cite{liu2026voyagemoe}. Existing evidence is therefore narrow and mixed, leaving it unclear whether recent pretrained MoE LLMs can be competitive first-stage retrievers.


This motivates our central question: \emph{can pretrained MoE LLMs outperform dense retrievers at comparable active parameter counts while matching larger dense retrievers at lower query encoding cost?} The answer does not directly follow from generative-model results. Retrieval requires encoding an entire sequence into a single representation rather than predicting the next token, so routing patterns learned during language-model pretraining may not transfer directly to retrieval. Moreover, active parameter count alone does not determine query encoding cost, which also depends on model depth, how many experts are activated across a batch, and the serving configuration \cite{chu2025dynaexq}.

To answer this question, we conduct a controlled study of MoE backbones for retrieval. We train MoE and dense LLMs from several families using the same training procedure, compare each MoE retriever on BEIR \cite{thakur2021beir} with a dense model from the same family with a comparable active parameter count (its active-matched control), and measure query and document encoding time under the same hardware and serving configuration. Across model families, MoE retrievers outperform their active-matched controls and can match larger dense retrievers at lower query encoding cost. Figure~\ref{fig:frontier} summarizes this effectiveness--efficiency trade-off. Our contributions are as follows:

\begin{itemize}
\item We present a controlled comparison of MoE and dense LLM backbones as first-stage retrievers. Across several model families, MoE retrievers outperform dense counterparts with comparable active parameter counts by 1.4--3.0 nDCG@10 points on BEIR-13.

\item We show that MoE retrievers can match larger dense retrievers at lower encoding cost. Qwen3-MoE matches Qwen3-8B with 59\% fewer active parameters and 18\% lower query encoding time.

\item We show that the number of experts used for query encoding can be reduced without retraining or re-indexing. With half as many experts per query token, an MoE retriever activates 19\% fewer
  parameters and reduces query encoding time by 26\%, while retaining more than 99\% of its effectiveness.

  \item We show that recent rerankers yield limited gains over strong MoE first stages. Reranking the top 200 adds at most 2.2 nDCG@10 points, while the two strongest MoE retrievers without reranking match or exceed every reranked configuration we evaluate.
\end{itemize}
Together, these results show that pretrained MoE LLMs can provide a strong and efficient alternative to dense LLMs for first-stage retrieval.

\section{Related Work}

\paragraph{LLM-based dense retrieval.}
RepLLaMA established a practical recipe for adapting decoder-only LLMs into
bi-encoders \cite{ma2024repllama}, and subsequent work explored LLM-generated
training data \cite{wang2024e5mistral}, bidirectional attention
\cite{behnamghader2024llm2vec}, latent-attention pooling
\cite{lee2025nvembed}, and multi-stage training
\cite{zhang2025qwen3embedding}. Across GTR and Qwen3 Embedding, larger
backbones generally yield stronger dense retrievers
\cite{ni2022gtr,zhang2025qwen3embedding}. A parallel line of work reduces
query encoding cost through distillation or structural compression of dense
retrievers
\cite{wang2023distill,cohen2024extremely,ma2025lightretriever,lei2025effir}.
EffiR, for example, reduces model depth and MLP width before retrieval
fine-tuning \cite{lei2025effir}. We study a complementary source of efficiency
already present in pretrained MoE backbones, comparing them with dense
backbones under the same training procedure and serving configuration.

\paragraph{MoE backbones for retrieval.}
Existing work on MoE architectures for retrieval spans several distinct settings. Nomic Embed v2 trains a purpose-built, BERT-scale MoE encoder for text embedding \cite{nussbaum2025nomic}, while GritLM evaluates a single Mixtral-based model under a joint generative and embedding objective and reports lower MTEB retrieval effectiveness than its dense 7B counterpart \cite{muennighoff2025gritlm}. Voyage AI has released proprietary MoE embedding models, but provides few public details about their architectures or measured efficiency \cite{liu2026voyagemoe}. MoEE \cite{li2025moee} instead uses the routing weights of a frozen MoE, alone or together
with its hidden states, to construct training-free embeddings without
fine-tuning the backbone for retrieval. Concurrent work on
Qwen3 MoE studies a cross-encoder reranker rather than first-stage retrieval
\cite{xu2026tevatron}. Other work uses ``mixture of experts'' for added expert modules, task- or domain-specific experts, routed adapters, or mixtures of whole retrievers \cite{sokli2025sbmoe,cheng2023taser,romero2025mote,kasela2024desireme,lee2025routerretriever,kalra2025mor}; these approaches introduce or compose expert components for retrieval rather than studying pretrained MoE LLM backbones directly as first-stage retrievers. We instead study pretrained MoE LLMs from five families, with controlled comparisons against active-matched dense models within three families under the same training procedure and serving configuration.


\paragraph{Inference-time expert reduction.}
Prior work on generative MoE models has explored reducing expert computation
after training through expert skipping
\cite{lu2024notallexperts,zhong2024adapmoe} or budget allocation
\cite{liu2026allocmoe}. Retrieval creates a different asymmetric setting
because queries and documents are encoded independently: the number of experts
used for query encoding can be reduced while leaving the corpus index
unchanged. We study this query-side reduction across multiple pretrained MoE
backbones and measure its effect on both retrieval effectiveness and query
encoding time.

\section{Training LLM Retrievers}
  \label{sec:method}

  Given a query $Q$ and a corpus of documents $\mathcal{C}$, first-stage retrieval aims to rank documents by their relevance to $Q$ and return the highest-scoring candidates. We follow the standard
  dense bi-encoder formulation, in which queries and documents are encoded independently and ranked by the similarity of their vector representations \cite{karpukhin2020dpr}.

  \subsection{Adapting LLMs for Retrieval}
\label{sec:adapt}

Dense retrievers have traditionally been built on bidirectional encoders such as BERT, which attend over the whole input and use the representation of a prepended \texttt{[CLS]} token as the embedding \cite{devlin2019bert,karpukhin2020dpr}. Decoder-only LLMs are instead unidirectional: each token attends only to the tokens before it, so a prepended token would see none of the input, and only the final token has seen all of it. Following RepLLaMA \cite{ma2024repllama}, we therefore append an end-of-sequence token to each query and document and take the final-layer hidden state of that token as the representation of the whole input,
\begin{equation*}
\mathbf{V}_X = \mathrm{Decoder}(X\,\langle\mathrm{eos}\rangle)[-1],
\end{equation*}
where $\mathrm{Decoder}(\cdot)$ returns the final-layer representation of each input token and $[-1]$ selects the end-of-sequence token. Here, $X$ can be either a query $Q$ or a document $D$. We L2-normalize the representations and score a document against a query by the dot product $\mathrm{Sim}(Q,D)=\mathbf{V}_Q^{\top}\mathbf{V}_D$.

\subsection{Training Objective}
\label{sec:objective}

Given a query $Q$, a relevant document $D^{+}$, and a set of negative documents $\{D_i^{-}\}$, we train the retriever with the InfoNCE loss
\begin{multline*}
\mathcal{L}(Q,D^{+},\{D_i^{-}\}) = -\log p(D=D^{+}\mid Q) \\
= -\log
\frac{e^{\mathrm{Sim}(Q,D^{+})/\tau}}
{e^{\mathrm{Sim}(Q,D^{+})/\tau} + \sum_{i} e^{\mathrm{Sim}(Q,D_i^{-})/\tau}}.
\end{multline*}
Here, $\tau$ is a temperature parameter. The negatives combine hard negatives, which are non-relevant documents paired with $Q$ in the training collection, and in-batch negatives, which are documents associated with the other queries in the same batch \cite{karpukhin2020dpr,ma2024repllama}. We use the same bi-encoder formulation and training objective for all dense and MoE backbones.

\section{Experimental Setup}
\label{sec:setup}
\paragraph{Model Backbones.}
We study MoE and dense decoder-only LLMs from five families: OLMo \cite{groeneveld2024olmo,muennighoff2025olmoe}, LFM2 \cite{liquid2025lfm2}, Qwen3 \cite{yang2025qwen3}, DeepSeek \cite{deepseek2024llm,deepseek2024v2}, and Gemma \cite{gemma2025gemma3,gemma2026gemma4}. Our primary comparisons pair OLMoE-1B-7B, LFM2-8B-A1B, and Qwen3-30B-A3B with dense models from the same family having similar active parameter counts, which we call active-matched dense controls. We also include larger dense models from these families as reference points. DeepSeek-V2-Lite and Gemma-4 26B-A4B are additional MoE backbones, but their dense comparisons differ in model generation or architecture and are therefore treated as descriptive rather than controlled. Further model details are provided in Appendix~\ref{app:backbones}.

\paragraph{Training Data.}
  We train all retrievers on the recently released RLHN dataset \cite{thakur2025rlhn}. RLHN is a lite version of the larger BGE retrieval training mixture \cite{li2025bgeicl} and uses LLM judgments
  to identify and relabel false hard negatives, yielding cleaner training supervision. It has been reported to improve retrieval effectiveness while reducing training cost
\cite{thakur2025rlhn,xu2026laconic}.

\paragraph{Implementation Details.}
We train all retrievers in Tevatron \cite{gao2023tevatron} using LoRA \cite{hu2022lora}. For MoE backbones, the adapters cover the expert parameters while the
pretrained routers remain frozen, as keeping them frozen was more effective
in preliminary experiments (see Appendix~\ref{app:router}). Each query is paired with one relevant document and 15 hard negatives; with a global batch size of 32, each query is contrasted against 512 candidate documents. Queries and documents are truncated to 192 tokens during training, and we use temperature $\tau=0.01$, bf16 precision, and gradient checkpointing. Full training configurations are provided in Appendix~\ref{app:training}.

\paragraph{Evaluation.}
We evaluate with exact search on 13 BEIR datasets \cite{thakur2021beir} and report nDCG@10. Queries and documents are truncated to 512 tokens at evaluation time.

\section{Are MoE Backbones Effective Retrievers?}
\label{sec:effectiveness}

\begin{table*}[t]
\caption{BEIR-13 effectiveness (nDCG@10) of dense and MoE retrievers trained with the same procedure, grouped by model family; $\dagger$ marks the MoE retrievers. \emph{Active} is the active parameter count in billions. Bold marks the best average within each family; for each MoE retriever, the parentheses give its relative improvement over the family's active-matched dense control and larger dense model, in that order, with -- where the family has no such model.}

\label{tab:main}
\centering\footnotesize\setlength{\tabcolsep}{2pt}\renewcommand{\arraystretch}{1.15}
\resizebox{\textwidth}{!}{\begin{tabular}{@{}l l @{\hspace{3pt}} l @{\hspace{14pt}} rrrrrrrrrrrrr @{\hspace{4pt}} | @{\hspace{4pt}} l@{}}
\toprule
\textbf{Family} & \textbf{Model} & \textbf{Active} & \textbf{Arg} & \textbf{Clm} & \textbf{DBP} & \textbf{FEV} & \textbf{FiQ} & \textbf{Hot} & \textbf{NFC} & \textbf{NQ} & \textbf{Quo} & \textbf{SCI} & \textbf{ScF} & \textbf{TRC} & \textbf{Tou} & \textbf{Average} \\
\midrule
 & OLMo-1B & 1.28 & 57.9 & 37.5 & 42.5 & 88.0 & 44.3 & 77.1 & 37.2 & 62.1 & 83.8 & 24.3 & 75.5 & 80.0 & 26.2 & 56.6 \\
OLMo & OLMo-MoE-A1B$^{\dagger}$ & 1.28 & 74.9 & 38.3 & 46.2 & 89.7 & 48.4 & 79.6 & 41.1 & 66.7 & 82.6 & 26.1 & 76.3 & 80.4 & 24.7 & \textbf{59.6} {\fontsize{6}{7}\selectfont(\textcolor{green!55!black}{+5.3\%}, \textcolor{green!55!black}{+1.7\%})} \\
 & OLMo-7B & 6.89 & 61.7 & 36.4 & 46.8 & 89.0 & 56.1 & 81.9 & 39.4 & 65.9 & 87.4 & 27.8 & 77.5 & 68.6 & 23.8 & 58.6 \\
\midrule
 & LFM2-1.2B & 1.17 & 67.5 & 36.5 & 44.6 & 87.2 & 43.9 & 74.9 & 38.9 & 62.6 & 80.2 & 24.3 & 72.9 & 84.6 & 31.8 & 57.7 \\
LFM2 & LFM2-MoE-A1B$^{\dagger}$ & 1.56 & 73.9 & 39.8 & 46.5 & 87.5 & 46.3 & 77.3 & 40.6 & 66.4 & 78.2 & 26.7 & 77.5 & 82.0 & 29.1 & \textbf{59.4} {\fontsize{6}{7}\selectfont(\textcolor{green!55!black}{+2.9\%}, \textcolor{green!55!black}{+1.9\%})} \\
 & LFM2-2.6B & 2.57 & 69.0 & 39.1 & 44.3 & 86.1 & 46.7 & 76.9 & 39.8 & 65.8 & 76.6 & 26.5 & 74.6 & 84.5 & 27.4 & 58.3 \\
\midrule
 & Qwen3-4B & 4.02 & 74.3 & 42.4 & 48.0 & 89.4 & 52.3 & 80.6 & 39.3 & 67.7 & 83.5 & 28.8 & 74.4 & 83.9 & 27.2 & 60.9 \\
Qwen3 & Qwen3-MoE-A3B$^{\dagger}$ & 3.35 & 78.7 & 41.8 & 48.6 & 90.4 & 57.1 & 83.7 & 42.0 & 70.8 & 80.5 & 32.2 & 77.3 & 83.6 & 23.4 & 62.3 {\fontsize{6}{7}\selectfont(\textcolor{green!55!black}{+2.3\%}, \textcolor{red!75!black}{-0.2\%})} \\
 & Qwen3-8B & 8.19 & 74.3 & 41.7 & 51.6 & 91.3 & 58.3 & 83.1 & 40.9 & 70.3 & 81.3 & 29.7 & 75.0 & 87.3 & 26.7 & \textbf{62.4} \\
\midrule
DeepSeek & DeepSeek-MoE-A2B$^{\dagger}$ & 2.66 & 75.0 & 42.3 & 47.2 & 89.7 & 53.3 & 81.7 & 41.5 & 67.2 & 85.1 & 27.3 & 76.6 & 79.0 & 21.5 & \textbf{60.6} {\fontsize{6}{7}\selectfont(--, \textcolor{green!55!black}{+1.2\%})} \\
 & DeepSeek-7B & 6.91 & 70.6 & 41.4 & 47.0 & 89.4 & 49.7 & 78.4 & 39.4 & 65.4 & 83.9 & 25.4 & 74.7 & 84.6 & 28.4 & 59.9 \\
\midrule
Gemma & Gemma-3-4B & 3.88 & 78.0 & 41.6 & 48.7 & 89.0 & 51.6 & 81.4 & 41.7 & 68.0 & 82.5 & 27.5 & 74.8 & 80.2 & 22.9 & 60.6 \\
 & Gemma-4-MoE-A4B$^{\dagger}$ & 3.82 & 82.0 & 43.0 & 49.2 & 90.8 & 55.4 & 85.0 & 42.1 & 70.4 & 85.0 & 30.4 & 78.8 & 79.5 & 20.3 & \textbf{62.5} {\fontsize{6}{7}\selectfont(\textcolor{green!55!black}{+3.0\%}, --)} \\
\bottomrule
\end{tabular}
}
\end{table*}

We first examine whether pretrained MoE backbones are effective for first-stage retrieval. Table~\ref{tab:main} reports nDCG@10 on 13 BEIR datasets, with retrievers grouped by model family. For each controlled family, we compare the MoE retriever with an active-matched dense control and a larger dense model from the same family, using the training procedure described in Sections~\ref{sec:method} and \ref{sec:setup}.

\paragraph{MoE Retrievers Outperform Active-Matched Dense Controls}
\label{sec:active-matched}

Across all three controlled families, the MoE retriever outperforms its
active-matched dense control. OLMo-MoE improves the BEIR-13 average from 56.6
to 59.6 nDCG@10, a gain of 3.0 points (5.3\% relative). LFM2-MoE improves
over LFM2-1.2B by 1.7 points (2.9\%), while Qwen3-MoE improves over Qwen3-4B
by 1.4 points (2.3\%). Gemma shows a similar gain at nearly identical active
parameter counts, with Gemma-4-MoE reaching 62.5 compared with 60.6 for
Gemma-3-4B (3.0\% relative).

The improvements are also consistent across datasets. The MoE retriever
outperforms its active-matched dense control on 11 of 13 datasets for OLMo,
10 of 13 for LFM2, and 9 of 13 for Qwen3. Across these comparisons, the gains
occur at comparable active parameter counts: OLMo-1B and OLMo-MoE both
activate 1.28B parameters, LFM2-MoE activates 1.56B compared with 1.17B for
LFM2-1.2B, Qwen3-MoE activates 3.35B compared with 4.02B for Qwen3-4B, and
the two Gemma models activate 3.82B and 3.88B.

\paragraph{MoE Retrievers Are Competitive with Larger Dense Models}
\label{sec:larger}

The MoE retrievers also match or exceed larger dense models from the same
families while activating substantially fewer parameters. OLMo-MoE exceeds
OLMo-7B by 1.0 point (59.6 vs.\ 58.6) with 5.4$\times$ fewer active
parameters, while LFM2-MoE exceeds LFM2-2.6B by 1.1 points (59.4 vs.\ 58.3)
with 1.6$\times$ fewer. Qwen3-MoE reaches 62.3, essentially matching Qwen3-8B
at 62.4 with 59\% fewer active parameters.

The descriptive DeepSeek comparison shows the same pattern: DeepSeek-MoE
exceeds DeepSeek-7B by 0.7 points (60.6 vs.\ 59.9) with 2.6$\times$ fewer
active parameters. Gemma-4-MoE is the strongest retriever overall at 62.5
nDCG@10. Overall, the MoE retrievers match or exceed dense models that have
1.6--5.4$\times$ as many active parameters. We next examine whether these
reductions in active parameters translate into lower query encoding cost.

\section{Are MoE Retrievers Efficient?}
\label{sec:efficiency}

Section~\ref{sec:effectiveness} showed that MoE retrievers can match or exceed larger dense retrievers while activating substantially fewer parameters. Active parameter count, however, does not
directly determine inference cost, which also depends on model architecture and the serving configuration. We therefore measure both query and document encoding time directly.

We measure query encoding time on 1,000 queries from the Natural Questions (NQ) dataset in BEIR at batch size 128, and document encoding time on 1,000 documents from the NQ corpus at batch size
64, both on a single NVIDIA H100 GPU. Models are evaluated in bf16 using Hugging Face Transformers with \texttt{torch.compile}, with MoE retrievers using their trained number of experts.
  Table~\ref{tab:efficiency} reports encoding time in microseconds per query and document; additional timing and serving details are provided in Appendix~\ref{app:efficiency}.

\begin{table}[t]
\caption{Query and document encoding time on a single NVIDIA H100 GPU, with the BEIR-13 nDCG@10 average of Table~\ref{tab:main} for reference. MoE rows ($\dagger$) use their trained expert count. \emph{Active} is the active parameter count in billions. Bold denotes the highest nDCG@10 and the lowest time within each family. Parentheses compare each MoE retriever with its active-matched dense control and larger dense reference, respectively ($\uparrow$ faster, $\downarrow$ slower).}
\label{tab:efficiency}
\centering\footnotesize\setlength{\tabcolsep}{2pt}\renewcommand{\arraystretch}{1.15}
\resizebox{\columnwidth}{!}{\begin{tabular}{@{}l @{\hspace{3pt}} l @{\hspace{6pt}} l @{\hspace{8pt}} l l@{}}
\toprule
\textbf{Model} & \textbf{Active} & \textbf{nDCG@10} & \textbf{$\mu$s/q} & \textbf{$\mu$s/d} \\
\midrule
OLMo-1B & 1.28 & 56.6 & \textbf{201} & \textbf{2,455} \\
OLMo-MoE-A1B$^{\dagger}$ & 1.28 & \textbf{59.6} & 316 {\fontsize{6}{7}\selectfont(\textcolor{red!75!black}{$\downarrow$1.6$\times$}, \textcolor{green!55!black}{$\uparrow$3.7$\times$})} & 3,451 {\fontsize{6}{7}\selectfont(\textcolor{red!75!black}{$\downarrow$1.4$\times$}, \textcolor{green!55!black}{$\uparrow$4.1$\times$})} \\
OLMo-7B & 6.89 & 58.6 & 1,168 & 14,104 \\
\midrule
LFM2-1.2B & 1.17 & 57.7 & \textbf{204} & \textbf{2,350} \\
LFM2-MoE-A1B$^{\dagger}$ & 1.56 & \textbf{59.4} & 374 {\fontsize{6}{7}\selectfont(\textcolor{red!75!black}{$\downarrow$1.8$\times$}, \textcolor{green!55!black}{$\uparrow$1.2$\times$})} & 3,837 {\fontsize{6}{7}\selectfont(\textcolor{red!75!black}{$\downarrow$1.6$\times$}, \textcolor{green!55!black}{$\uparrow$1.4$\times$})} \\
LFM2-2.6B & 2.57 & 58.3 & 460 & 5,330 \\
\midrule
Qwen3-4B & 4.02 & 60.9 & \textbf{702} & \textbf{8,873} \\
Qwen3-MoE-A3B$^{\dagger}$ & 3.35 & 62.3 & 1,037 {\fontsize{6}{7}\selectfont(\textcolor{red!75!black}{$\downarrow$1.5$\times$}, \textcolor{green!55!black}{$\uparrow$1.2$\times$})} & 11,050 {\fontsize{6}{7}\selectfont(\textcolor{red!75!black}{$\downarrow$1.2$\times$}, \textcolor{green!55!black}{$\uparrow$1.4$\times$})} \\
Qwen3-8B & 8.19 & \textbf{62.4} & 1,271 & 15,798 \\
\midrule
DeepSeek-MoE-A2B$^{\dagger}$ & 2.66 & \textbf{60.6} & \textbf{587} {\fontsize{6}{7}\selectfont(--, \textcolor{green!55!black}{$\uparrow$1.8$\times$})} & \textbf{6,402} {\fontsize{6}{7}\selectfont(--, \textcolor{green!55!black}{$\uparrow$2.2$\times$})} \\
DeepSeek-7B & 6.91 & 59.9 & 1,078 & 14,124 \\
\midrule
Gemma-3-4B & 3.88 & 60.6 & \textbf{619} & \textbf{7,117} \\
Gemma-4-MoE-A4B$^{\dagger}$ & 3.82 & \textbf{62.5} & 1,115 {\fontsize{6}{7}\selectfont(\textcolor{red!75!black}{$\downarrow$1.8$\times$}, --)} & 12,658 {\fontsize{6}{7}\selectfont(\textcolor{red!75!black}{$\downarrow$1.8$\times$}, --)} \\
\bottomrule
\end{tabular}
}
\end{table}

\begin{figure*}[t]
  \centering
  \includegraphics[width=\textwidth]{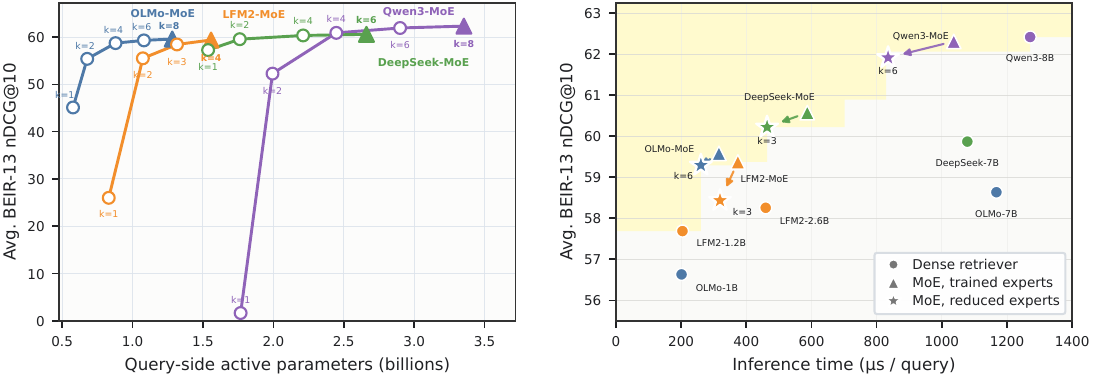}
  \caption{Query-side expert reduction.
(a) Average BEIR-13 nDCG@10 against active query parameters as the query
expert count $k$ is reduced below the trained count ($\blacktriangle$) with
the corpus index fixed.
(b) The effectiveness--efficiency view of Figure~\ref{fig:frontier}, with each
MoE retriever at its trained expert count ($\blacktriangle$) and selected
reduced query expert count ($\star$), joined by an arrow; dense retrievers
($\bullet$) and the Pareto frontier (shaded) are as in
Figure~\ref{fig:frontier}.}
  \label{fig:topk}
\end{figure*}

\paragraph{MoE Retrievers Are Cheaper Than the Larger Dense Models They Match.}
The MoE retrievers consistently require less encoding time than the larger dense models with comparable effectiveness. OLMo-MoE reduces query encoding time from 1,168 to 316 $\mu$s, a 3.7$\times$ reduction, and document encoding time from 14,104 to 3,451 $\mu$s, a 4.1$\times$ reduction, relative to OLMo-7B. LFM2-MoE is 1.2$\times$ faster for query encoding and 1.4$\times$ faster for document encoding than LFM2-2.6B. Qwen3-MoE, which essentially matches Qwen3-8B in retrieval effectiveness, reduces query encoding time from 1,271 to 1,037 $\mu$s and document encoding time from 15,798 to 11,050 $\mu$s. The descriptive DeepSeek comparison shows the same pattern, with DeepSeek-MoE reducing query and document encoding time by 1.8$\times$ and 2.2$\times$, respectively.

MoE retrievers are, however, slower than their active-matched dense controls while achieving higher retrieval effectiveness. Across OLMo, LFM2, and Qwen3, the MoE models require 1.5--1.8$\times$ more query encoding time despite having comparable active parameter counts, and document encoding shows the same broad pattern. Gemma similarly requires more encoding time than its active-scale dense comparison. These results reinforce that active parameter count alone is not a reliable proxy for runtime. Instead, the efficiency advantage of MoE backbones emerges when comparing models at similar retrieval effectiveness: they can match substantially larger dense retrievers at lower query and document encoding cost.

Figure~\ref{fig:frontier} summarizes the overall effectiveness--efficiency trade-off and also includes selected MoE retrievers at reduced query expert counts. As we show next, reducing the number of query-side experts can further lower encoding cost without retraining or re-indexing, while largely preserving retrieval effectiveness.

\section{Can MoE Retrievers Use Fewer Experts at Query Time?}
\label{sec:experts}

Section~\ref{sec:efficiency} showed that MoE retrievers can match larger dense
retrievers at lower encoding cost, while remaining slower than their
active-matched dense controls. We therefore ask whether query encoding can be
made cheaper still. MoE backbones are trained with a fixed number of experts
selected per token, which we call the trained expert count. Because queries
and documents are encoded independently, we can reduce the number of experts
used for query encoding after training while leaving the corpus index
unchanged. This requires neither retraining nor re-indexing.

For each MoE retriever, we re-encode only the queries with a smaller expert
count $k$ and search the unchanged corpus index built at the trained count.
Query encoding time is measured as in Section~\ref{sec:efficiency}.
Figure~\ref{fig:topk}(a) shows how BEIR-13 effectiveness changes as $k$ is
reduced for OLMo-MoE, LFM2-MoE, Qwen3-MoE, and DeepSeek-MoE, while
Figure~\ref{fig:topk}(b) places selected reduced-expert settings in the
effectiveness--efficiency plane of Figure~\ref{fig:frontier}. Detailed per-dataset results and query encoding times for additional query
expert counts, including Gemma-4-MoE, are provided in
Appendix~\ref{app:sweeps}.

\paragraph{MoE Retrievers Preserve Effectiveness with Fewer Query Experts at Lower Cost.}

\begin{table}[htbp]
\caption{Query-side expert reduction. We report the smallest query expert
count retaining at least 99\% of trained-count BEIR-13 nDCG@10, except for
LFM2-MoE, for which we use one fewer expert. Reductions in active query
parameters and query encoding time are relative to the trained count.}
\label{tab:topk}
\centering\footnotesize
\resizebox{\columnwidth}{!}{\begin{tabular}{@{}l c r r r@{}}
\toprule
Model & $k$ & Active $\downarrow$ & nDCG@10 retained & Query time $\downarrow$ \\
\midrule
OLMo-MoE-A1B & 8 $\to$ 6 & 15.7\% & 99.5\% & 17.5\% \\
LFM2-MoE-A1B & 4 $\to$ 3 & 15.5\% & 98.4\% & 14.7\% \\
Qwen3-MoE-A3B & 8 $\to$ 6 & 13.5\% & 99.4\% & 19.4\% \\
DeepSeek-MoE-A2B & 6 $\to$ 3 & 25.4\% & 99.4\% & 21.0\% \\
Gemma-4-MoE-A4B & 8 $\to$ 4 & 18.7\% & 99.2\% & 26.2\% \\
\bottomrule
\end{tabular}
}
\end{table}

Table~\ref{tab:topk} reports, for each MoE retriever, the smallest query expert
count that retains at least 99\% of its trained-count effectiveness, except
LFM2-MoE, for which we use one fewer expert than its trained count of four.
OLMo-MoE, Qwen3-MoE, DeepSeek-MoE, and Gemma-4-MoE retain 99.2--99.5\% of
their BEIR-13 nDCG@10 while activating 13.5--25.4\% fewer parameters per
query and reducing query encoding time by 17.5--26.2\%. LFM2-MoE retains
98.4\% of its effectiveness while reducing query encoding time by 14.7\%.
Figure~\ref{fig:topk}(a) shows that effectiveness generally degrades gradually
under moderate reductions in $k$, with sharper losses only at the smallest
expert counts.

Figure~\ref{fig:topk}(b) places these settings on the
effectiveness--efficiency plane of Figure~\ref{fig:frontier}. The reduced
settings move toward lower query encoding time with little loss in
effectiveness, narrowing the efficiency gap to the active-matched dense
controls while remaining competitive with the larger dense models. OLMo-MoE
at $k=6$ reaches 59.3 nDCG@10 at 261 $\mu$s per query, compared with 58.6 at
1,168 $\mu$s for OLMo-7B, a 4.5$\times$ reduction in query encoding time.
Qwen3-MoE at $k=6$ remains within 0.5 points of Qwen3-8B while reducing query
encoding time from 1,271 to 835 $\mu$s. Because only query-side routing is
changed, these savings require neither retraining nor rebuilding the corpus
index.

\section{Do Strong MoE Retrievers Still Benefit from Reranking?}
\label{sec:reranking}

Two-stage retrieval pipelines commonly use a fast first-stage retriever to
generate a candidate set and a more expensive reranker to refine the ranking.
We ask whether such reranking still provides meaningful gains when the
first-stage MoE retriever is already strong. For each MoE retriever, we rerank
its top 200 documents with three released pointwise rerankers:
mxbai-rerank-large-v2 \cite{li2025prorank}, Qwen3-Reranker-0.6B
\cite{zhang2025qwen3embedding}, and the older RankLLaMA-v1-7B
\cite{ma2024repllama}, and report nDCG@10 in Table~\ref{tab:reranking}.

\begin{table}[htbp]
\caption{Reranking the top 200 documents of each MoE first stage with three released pointwise rerankers (BEIR-13 nDCG@10). \emph{None} is the unreranked first stage; parentheses give the relative change from it.}
\label{tab:reranking}
\centering\small\setlength{\tabcolsep}{3pt}
\resizebox{\columnwidth}{!}{\begin{tabular}{@{}l r rrr@{}}
\toprule
First stage & None & mxbai-v2 & Qwen3-0.6B & RankLLaMA-v1-7B \\
\midrule
LFM2-MoE-A1B & 59.4 & 61.4 {\fontsize{6}{7}\selectfont(\textcolor{green!55!black}{+3.4\%})} & 60.4 {\fontsize{6}{7}\selectfont(\textcolor{green!55!black}{+1.7\%})} & 55.1 {\fontsize{6}{7}\selectfont(\textcolor{red!75!black}{-7.2\%})} \\
OLMo-MoE-A1B & 59.6 & 61.8 {\fontsize{6}{7}\selectfont(\textcolor{green!55!black}{+3.7\%})} & 60.3 {\fontsize{6}{7}\selectfont(\textcolor{green!55!black}{+1.2\%})} & 55.5 {\fontsize{6}{7}\selectfont(\textcolor{red!75!black}{-7.0\%})} \\
DeepSeek-MoE-A2B & 60.6 & 61.5 {\fontsize{6}{7}\selectfont(\textcolor{green!55!black}{+1.5\%})} & 60.4 {\fontsize{6}{7}\selectfont(\textcolor{red!75!black}{-0.3\%})} & 55.4 {\fontsize{6}{7}\selectfont(\textcolor{red!75!black}{-8.5\%})} \\
Qwen3-MoE-A3B & 62.3 & 61.9 {\fontsize{6}{7}\selectfont(\textcolor{red!75!black}{-0.6\%})} & 60.6 {\fontsize{6}{7}\selectfont(\textcolor{red!75!black}{-2.7\%})} & 55.5 {\fontsize{6}{7}\selectfont(\textcolor{red!75!black}{-10.9\%})} \\
Gemma-4-MoE-A4B & 62.5 & 61.4 {\fontsize{6}{7}\selectfont(\textcolor{red!75!black}{-1.6\%})} & 60.6 {\fontsize{6}{7}\selectfont(\textcolor{red!75!black}{-3.0\%})} & 55.1 {\fontsize{6}{7}\selectfont(\textcolor{red!75!black}{-11.8\%})} \\
\bottomrule
\end{tabular}
}
\end{table}

\paragraph{Reranking Provides Limited Gains over Strong MoE First Stages.}
The benefit of reranking decreases as the first-stage retriever becomes
stronger. mxbai-rerank-large-v2 improves the three weaker MoE first stages by
0.9--2.2 nDCG@10 points, with the largest gain on OLMo-MoE, but reduces
Qwen3-MoE and Gemma-4-MoE by 0.4 and 1.0 points, respectively.
Qwen3-Reranker-0.6B provides smaller gains, adding at most 1.0 point and also
reducing the effectiveness of the two strongest first stages. The older
RankLLaMA-v1-7B reduces effectiveness for every MoE retriever we evaluate.

Most notably, the two strongest MoE first stages remain competitive with every
two-stage configuration without using a reranker: Qwen3-MoE and Gemma-4-MoE
reach 62.3 and 62.5 nDCG@10, respectively, matching or exceeding the best
reranked result. These results do not imply that reranking is generally
ineffective; they are limited to three released pointwise rerankers and a
reranking depth of 200. Rather, they show that as the first-stage retriever
becomes stronger, the additional effectiveness provided by reranking can
become small or even negative.

\section{Discussion}

Our results show that active parameter count alone does not explain retrieval
effectiveness for MoE backbones. Across several model families, MoE retrievers
outperform dense controls with similar active parameter counts, suggesting that
the larger pretrained capacity of MoE models remains useful after retrieval
fine-tuning. These comparisons should not be interpreted as isolating sparse
routing itself as the cause of the gains, since the MoE and dense backbones
also differ in total model capacity. Nevertheless, the consistent gains across
multiple families indicate that the result is not specific to a single
architecture.

Active parameter count is also not a direct measure of inference cost. The MoE
retrievers are slower than their active-matched dense controls, but can be
substantially faster than larger dense retrievers with similar effectiveness.
Runtime therefore depends on the architecture and serving configuration in
addition to the number of active parameters. Query-side expert reduction
provides another way to lower online encoding cost without retraining or
re-indexing, although the best expert count depends on the backbone and does
not reduce the memory required to store the full model. Finally, our reranking
results suggest that stronger first-stage retrievers can reduce the benefit of
a second stage, although this result is limited to the rerankers and candidate
depth evaluated here.

\section{Conclusion}

We systematically studied pretrained MoE LLMs as first-stage retrievers.
Across several model families, MoE retrievers outperform active-matched dense
controls and can match substantially larger dense retrievers at lower encoding
cost. We further show that the query expert count can be reduced after training
without re-indexing, preserving nearly all retrieval effectiveness while
lowering query encoding cost. Recent pointwise rerankers provide limited gains
over the strongest MoE first stages and can even reduce their effectiveness.
Together, these results show that pretrained MoE backbones are a strong and
efficient alternative to dense LLMs for first-stage retrieval.

\section*{Limitations}

Our efficiency results focus on encoding cost rather than memory or index
footprint. MoE models have substantially larger total parameter counts than
their active parameter counts, although recent serving systems can reduce GPU
memory requirements through expert offloading and caching
\cite{xue2024moeinfinity,he2024expertflow,kamahori2024fiddler}. We do not
evaluate such memory-aware serving strategies. We also do not evaluate index
size or approximate nearest-neighbor search cost, which depend on embedding
dimensionality and index configuration rather than model active parameter
count and can differ across backbones.
The query-side expert reduction also depends on the backbone. Different MoE
models tolerate different reductions in the number of experts before retrieval
effectiveness degrades, so the reduced expert count must be selected separately
for each architecture.
Finally, our reranking study is limited to three released pointwise rerankers
at a candidate depth of 200. The results therefore show that these rerankers
provide limited gains over our strongest MoE first stages, but do not imply
that reranking more generally is unnecessary.


\bibliography{custom}

\onecolumn
\clearpage
\appendix

\section{Appendix}
\label{sec:appendix}

\subsection{Model Backbones}
\label{app:backbones}

Table~\ref{tab:backbones} summarizes the backbones used in our experiments,
including their released checkpoints, total and active parameter counts, and
the routing configuration of each MoE model. OLMo, LFM2, and Qwen3 form our
controlled comparisons: within each family, the MoE model is compared with a
dense model at a comparable active parameter count and with a larger dense
reference. DeepSeek and Gemma are treated as descriptive comparisons because
the available dense models differ in model generation or architecture.

DeepSeek-V2-Lite selects 6 of 64 routed experts per token in addition to two
shared experts. Gemma-3-4B is a multimodal checkpoint of which only the text
tower is used for retrieval, while Gemma-4-MoE applies a dense feed-forward
block alongside its routed experts; both are reflected in the reported active
parameter counts. Tied input and output embeddings are counted once.

Because we use the final hidden state directly as the retrieval embedding,
  the embedding dimensionality equals the backbone hidden size reported in
  Table~\ref{tab:backbones}. This dimension affects index storage and
  approximate nearest-neighbor search cost independently of the model's active
  parameter count: the MoE retrievers embed in 2,048 dimensions (2,816 for
  Gemma-4-MoE), whereas the 7B and 8B dense models embed in 4,096.
\begin{table}[htbp]
\caption{Backbones used in our experiments. \emph{Experts} gives the number
  of routed experts and the number selected per token. Shared or other
  always-active components are included in the reported active parameter counts.
  Parameter counts are in billions. \emph{Dim.} is the hidden size of the
  backbone, which equals the embedding dimension.}
\label{tab:backbones}
\centering\footnotesize\setlength{\tabcolsep}{5pt}
\begin{tabular}{@{}l l c r r r@{}}
\toprule
Model & Checkpoint & Experts (total / active) & Total (B) & Active (B) & Dim. \\
\midrule
OLMo-1B & \texttt{allenai/OLMo-1B-0724-hf} & -- & 1.28 & 1.28 & 2,048 \\
OLMo-MoE-A1B & \texttt{allenai/OLMoE-1B-7B-0125} & 64 / 8 & 6.92 & 1.28 & 2,048 \\
OLMo-7B & \texttt{allenai/OLMo-7B-0724-hf} & -- & 6.89 & 6.89 & 4,096 \\
\addlinespace[3pt]
LFM2-1.2B & \texttt{LiquidAI/LFM2-1.2B} & -- & 1.17 & 1.17 & 2,048 \\
LFM2-MoE-A1B & \texttt{LiquidAI/LFM2-8B-A1B} & 32 / 4 & 8.34 & 1.56 & 2,048 \\
LFM2-2.6B & \texttt{LiquidAI/LFM2-2.6B} & -- & 2.57 & 2.57 & 2,048 \\
\addlinespace[3pt]
Qwen3-4B & \texttt{Qwen/Qwen3-4B} & -- & 4.02 & 4.02 & 2,560 \\
Qwen3-MoE-A3B & \texttt{Qwen/Qwen3-30B-A3B} & 128 / 8 & 30.53 & 3.35 & 2,048 \\
Qwen3-8B & \texttt{Qwen/Qwen3-8B} & -- & 8.19 & 8.19 & 4,096 \\
\addlinespace[3pt]
DeepSeek-MoE-A2B & \texttt{deepseek-ai/DeepSeek-V2-Lite} & 64 / 6 (+2 shared) & 15.71 & 2.66 & 2,048 \\
DeepSeek-7B & \texttt{deepseek-ai/deepseek-llm-7b-base} & -- & 6.91 & 6.91 & 4,096 \\
\addlinespace[3pt]
Gemma-3-4B & \texttt{google/gemma-3-4b-pt} & -- & 4.30 & 3.88 & 2,560 \\
Gemma-4-MoE-A4B & \texttt{google/gemma-4-26B-A4B} & 128 / 8 & 25.81 & 3.82 & 2,816 \\
\bottomrule
\end{tabular}

\end{table}

\subsection{Training Configuration}
\label{app:training}
Tables~\ref{tab:training-common} and \ref{tab:training-model} summarize the
shared training configuration and model-specific LoRA settings, respectively.
All retrievers are trained on the RLHN dataset.\footnote{\url{https://huggingface.co/datasets/rlhn/rlhn-680K}}
LoRA adapters are applied to the query, key, value, and output projections in
attention layers and to the up, gate, and down projections in feed-forward
layers, including the experts of MoE layers. Embeddings, normalization layers,
and MoE router weights remain frozen. LoRA dropout is zero for all models.

\begin{table}[htbp]
\caption{Training configuration shared across all retrievers.}
\label{tab:training-common}
\centering
\small
\begin{tabular}{@{}lr@{}}
\toprule
Setting & Value \\
\midrule
Optimizer & AdamW \\
Learning rate & $1\times10^{-4}$ \\
Warmup & 5\% \\
Learning-rate schedule & Linear decay to zero \\
Maximum query length & 192 \\
Maximum document length & 192 \\
Hard negatives per query & 15 \\
Global batch size & 32 \\
Temperature $\tau$ & 0.01 \\
Precision & bf16 \\
Gradient checkpointing & Yes \\
\bottomrule
\end{tabular}
\end{table}

\begin{table}[htbp]
\caption{Model-specific LoRA settings and number of training epochs.}
\label{tab:training-model}
\centering
\small
\begin{tabular}{@{}lccc@{}}
\toprule
Model & LoRA $r$ & LoRA $\alpha$ & Epochs \\
\midrule
OLMo-1B          & 32 & 64 & 2 \\
OLMo-MoE-A1B     & 16 & 32 & 1 \\
OLMo-7B          & 16 & 32 & 1 \\
\addlinespace[2pt]
LFM2-1.2B        & 32 & 64 & 2 \\
LFM2-MoE-A1B     & 16 & 32 & 1 \\
LFM2-2.6B        & 32 & 64 & 2 \\
\addlinespace[2pt]
Qwen3-4B         & 32 & 64 & 2 \\
Qwen3-MoE-A3B    & 16 & 32 & 1 \\
Qwen3-8B         & 16 & 32 & 1 \\
\addlinespace[2pt]
DeepSeek-MoE-A2B & 16 & 32 & 1 \\
DeepSeek-7B      & 16 & 32 & 1 \\
\addlinespace[2pt]
Gemma-3-4B       & 32 & 64 & 2 \\
Gemma-4-MoE-A4B  & 16 & 32 & 1 \\
\bottomrule
\end{tabular}
\end{table}
\newpage
\subsection{Router Fine-Tuning}
\label{app:router}

In our main experiments, the pretrained MoE routers remain frozen during
retrieval fine-tuning. We additionally test updating the router weights
together with the LoRA adapters for OLMo-MoE, comparing two otherwise
identical runs with frozen and fine-tuned routers.

\begin{figure}[htbp]
    \centering
    \includegraphics[width=0.42\textwidth]{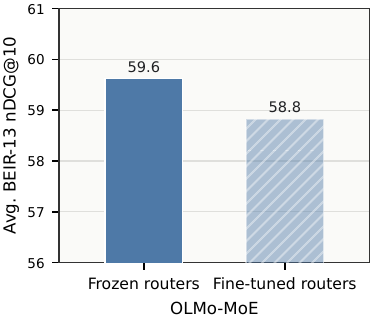}
    \caption{Effect of router fine-tuning for OLMo-MoE on BEIR-13 nDCG@10.
    The pretrained routers are either frozen or updated together with the
    LoRA adapters during retrieval fine-tuning.}
    \label{fig:router}
\end{figure}

In this ablation, fine-tuning the routers reduces the BEIR-13 average from
59.6 to 58.8 nDCG@10. Based on this result, we keep the pretrained routers
frozen for all MoE retrievers in our main experiments.
\newpage
\subsection{Expert-Count Sweeps}
\label{app:sweeps}

Table~\ref{tab:topkselected} reports per-dataset results for the query-side
expert-count settings considered in our analysis. For each MoE retriever, the
corpus index is built once using the trained expert count, and only the number
of experts used to encode queries is changed. We include the trained setting,
the reduced settings highlighted in Section~\ref{sec:experts}, and more
aggressive reductions to show how the effect of query-side expert reduction
varies across datasets.

The corresponding active query parameter counts, BEIR-13 averages, and query
encoding times are summarized in
Table~\ref{tab:topkselected-summary}. These results show that moderate
reductions in the query expert count generally preserve most retrieval
effectiveness, while more aggressive reductions lead to larger and more
model-dependent losses.

\begin{table}[htbp]
\caption{Per-dataset BEIR nDCG@10 for the five MoE retrievers at different
query expert counts $k$, with the BEIR-13 average (Avg.). The corpus index remains fixed at the trained expert
count, and only query encoding is changed.}
\label{tab:topkselected}
\centering
\scriptsize
\setlength{\tabcolsep}{2.5pt}
\resizebox{\textwidth}{!}{\begin{tabular}{@{}l l rrrrrrrrrrrrr r@{}}
\toprule
Model & $k$ & Arg & Clm & DBP & FEV & FiQ & Hot & NFC & NQ & Quo & SCI & ScF & TRC & Tou & Avg. \\
\midrule
OLMo-MoE-A1B & 8 (trained) & 74.8 & 38.3 & 46.1 & 89.7 & 48.3 & 79.6 & 41.0 & 66.7 & 82.6 & 26.2 & 76.3 & 80.7 & 24.4 & 59.6 \\
 & 6 & 73.7 & 38.6 & 46.1 & 89.6 & 48.3 & 79.1 & 40.8 & 66.6 & 82.7 & 25.9 & 76.3 & 79.8 & 23.4 & 59.3 \\
 & 4 & 71.9 & 38.3 & 45.2 & 88.7 & 47.0 & 77.5 & 40.6 & 65.4 & 82.7 & 25.6 & 77.4 & 78.1 & 24.9 & 58.7 \\
 & 1 & 56.6 & 29.6 & 34.7 & 70.1 & 25.3 & 50.5 & 34.4 & 43.6 & 77.5 & 17.1 & 67.3 & 65.0 & 14.8 & 45.1 \\
\addlinespace[2pt]
LFM2-MoE-A1B & 4 (trained) & 73.9 & 39.8 & 46.5 & 87.5 & 46.3 & 77.3 & 40.6 & 66.4 & 78.2 & 26.7 & 77.5 & 82.0 & 29.1 & 59.4 \\
 & 3 & 73.9 & 40.4 & 45.5 & 87.0 & 44.9 & 75.6 & 39.8 & 64.9 & 78.3 & 26.3 & 77.2 & 80.7 & 25.2 & 58.4 \\
 & 2 & 71.0 & 39.9 & 43.4 & 85.2 & 38.3 & 70.5 & 37.4 & 61.0 & 78.0 & 24.6 & 76.2 & 76.7 & 19.3 & 55.5 \\
 & 1 & 2.7 & 23.6 & 19.2 & 57.7 & 4.3 & 35.2 & 27.1 & 24.2 & 43.9 & 6.8 & 46.2 & 43.1 & 4.1 & 26.0 \\
\addlinespace[2pt]
Qwen3-MoE-A3B & 8 (trained) & 78.8 & 41.8 & 48.6 & 90.4 & 57.2 & 83.8 & 42.0 & 70.8 & 80.6 & 32.1 & 77.2 & 83.4 & 23.5 & 62.3 \\
 & 6 & 78.6 & 41.3 & 48.4 & 90.0 & 56.1 & 83.0 & 41.5 & 70.0 & 80.2 & 32.1 & 77.0 & 83.2 & 23.5 & 61.9 \\
 & 4 & 78.2 & 40.8 & 47.3 & 89.0 & 54.2 & 80.8 & 40.3 & 68.4 & 79.8 & 31.1 & 76.0 & 82.5 & 23.2 & 60.9 \\
 & 1 & 0.6 & 0.2 & 0.4 & 1.2 & 0.0 & 0.3 & 7.0 & 0.4 & 1.1 & 2.1 & 5.8 & 2.4 & 0.0 & 1.7 \\
\addlinespace[2pt]
DeepSeek-MoE-A2B & 6 (trained) & 75.0 & 42.3 & 47.2 & 89.7 & 53.3 & 81.7 & 41.5 & 67.2 & 85.1 & 27.3 & 76.6 & 79.0 & 21.5 & 60.6 \\
 & 4 & 74.2 & 41.8 & 47.0 & 89.8 & 52.8 & 81.4 & 41.9 & 67.0 & 85.2 & 27.0 & 76.5 & 78.6 & 21.3 & 60.3 \\
 & 3 & 73.5 & 41.3 & 47.2 & 89.8 & 52.3 & 80.8 & 41.7 & 66.6 & 85.1 & 26.8 & 76.1 & 79.6 & 22.0 & 60.2 \\
 & 1 & 68.0 & 38.8 & 43.5 & 85.6 & 50.0 & 74.5 & 38.9 & 63.3 & 85.0 & 24.8 & 74.9 & 74.6 & 22.3 & 57.3 \\
\addlinespace[2pt]
Gemma-4-MoE-A4B & 8 (trained) & 82.0 & 43.0 & 49.2 & 90.8 & 55.4 & 85.0 & 42.1 & 70.4 & 85.0 & 30.4 & 78.8 & 79.5 & 20.3 & 62.5 \\
 & 6 & 82.0 & 43.3 & 48.8 & 90.5 & 54.8 & 84.8 & 41.6 & 70.1 & 84.9 & 30.3 & 78.3 & 79.9 & 20.2 & 62.3 \\
 & 4 & 81.9 & 43.8 & 48.7 & 90.1 & 53.4 & 84.1 & 41.5 & 69.4 & 84.8 & 29.6 & 78.1 & 79.5 & 20.4 & 62.0 \\
 & 1 & 69.5 & 40.2 & 41.2 & 84.0 & 36.4 & 71.7 & 36.7 & 59.0 & 81.2 & 24.2 & 71.5 & 77.5 & 20.3 & 54.9 \\
\bottomrule
\end{tabular}
}
\end{table}

\begin{table}[htbp]
\caption{Active query parameters, BEIR-13 nDCG@10, and query encoding time for
the settings of Table~\ref{tab:topkselected}. Active $\downarrow$ and Query
time $\downarrow$ are reductions relative to the trained expert count, and
Retained is nDCG@10 relative to the trained count. Query encoding time is
measured on one H100 at batch size 128, as in Section~\ref{sec:efficiency}.}
\label{tab:topkselected-summary}
\centering
\footnotesize
\setlength{\tabcolsep}{5pt}
\begin{tabular}{@{}l l r r r r r r@{}}
\toprule
Model & $k$ & Active (B) & Active $\downarrow$ & nDCG@10 & Retained & $\mu$s/q & Query time $\downarrow$ \\
\midrule
OLMo-MoE-A1B & 8 (trained) & 1.28 & -- & 59.6 & 100.0\% & 316 & -- \\
 & 6 & 1.08 & 15.7\% & 59.3 & 99.5\% & 261 & 17.5\% \\
 & 4 & 0.88 & 31.4\% & 58.7 & 98.5\% & 216 & 31.6\% \\
 & 1 & 0.58 & 55.0\% & 45.1 & 75.7\% & 141 & 55.3\% \\
\addlinespace[2pt]
LFM2-MoE-A1B & 4 (trained) & 1.56 & -- & 59.4 & 100.0\% & 374 & -- \\
 & 3 & 1.32 & 15.5\% & 58.4 & 98.4\% & 319 & 14.7\% \\
 & 2 & 1.07 & 31.1\% & 55.5 & 93.5\% & 285 & 23.9\% \\
 & 1 & 0.83 & 46.7\% & 26.0 & 43.8\% & 228 & 39.0\% \\
\addlinespace[2pt]
Qwen3-MoE-A3B & 8 (trained) & 3.35 & -- & 62.3 & 100.0\% & 1,037 & -- \\
 & 6 & 2.90 & 13.5\% & 61.9 & 99.4\% & 835 & 19.4\% \\
 & 4 & 2.45 & 27.0\% & 60.9 & 97.7\% & 716 & 30.9\% \\
 & 1 & 1.77 & 47.3\% & 1.7 & 2.7\% & 510 & 50.9\% \\
\addlinespace[2pt]
DeepSeek-MoE-A2B & 6 (trained) & 2.66 & -- & 60.6 & 100.0\% & 587 & -- \\
 & 4 & 2.21 & 16.9\% & 60.3 & 99.6\% & 505 & 14.0\% \\
 & 3 & 1.99 & 25.4\% & 60.2 & 99.4\% & 464 & 21.0\% \\
 & 1 & 1.54 & 42.2\% & 57.3 & 94.5\% & 353 & 39.9\% \\
\addlinespace[2pt]
Gemma-4-MoE-A4B & 8 (trained) & 3.82 & -- & 62.5 & 100.0\% & 1,115 & -- \\
 & 6 & 3.47 & 9.3\% & 62.3 & 99.7\% & 934 & 16.2\% \\
 & 4 & 3.11 & 18.7\% & 62.0 & 99.2\% & 822 & 26.2\% \\
 & 1 & 2.57 & 32.7\% & 54.9 & 87.9\% & 618 & 44.5\% \\
\bottomrule
\end{tabular}

\end{table}
\newpage

\subsection{Efficiency Details}
\label{app:efficiency}

All encoding-time measurements use the same harness and serving configuration
described in Section~\ref{sec:efficiency}. We sample 1,000 queries and 1,000
documents once from the BEIR Natural Questions dataset using a fixed seed and
process them exactly as in retrieval evaluation, including the same input
format, appended end-of-sequence token, and truncation to 512 tokens.
Tokenization and data loading are excluded from the timed region.

Before timing, we perform one warm-up pass to trigger \texttt{torch.compile}
and other one-time initialization overheads. We then run five timed passes over
the full sample, synchronizing GPU execution before and after each pass, and
report the median. Query encoding is measured at batch size 128 and document
encoding at batch size 64. The reported microseconds per query and per document
reflect amortized batched encoding time.

\end{document}